\documentclass{article}
\usepackage[utf8]{inputenc}

\usepackage{url}            % simple URL typesetting
\usepackage{hyperref}
\usepackage{booktabs}       % professional-quality tables
\usepackage{amsfonts}       % blackboard math symbols
\usepackage{nicefrac}       % compact symbols for 1/2,
\usepackage{microtype}      % microtypography
\usepackage{xcolor}         % colors
\usepackage[pdftex]{graphicx}
\usepackage{subcaption}
\usepackage{wrapfig}
\usepackage{enumitem}
\usepackage{multirow}

\usepackage{bm}
\usepackage{bbm}
\usepackage{amssymb}
\usepackage{amsmath}
\usepackage{mathtools}
\usepackage[nameinlink]{cleveref}

\usepackage{steinmetz}

\hypersetup{
	colorlinks=true,
	linkcolor=myblue,
	citecolor=myblue,
	urlcolor=mygray,
}

\definecolor{myblue}{RGB}{0,0, 177}
\definecolor{mygray}{RGB}{103,103,106}

\newcommand{\R}{\mathbb{R}} %real numbers
\newcommand{\C}{\mathbb{C}} %complex numbers

\newcommand{\Y}{{\bf Y}}
\newcommand{\Source}{{\bf S}}

\newcommand{\x}{{\bf x}}
\newcommand{\y}{{\bf y}}
\newcommand{\s}{{\bf s}}
\newcommand{\n}{{\bf n}}

\usepackage[numbers]{natbib}

\usepackage{spconf}

\title{A Training Framework for Stereo-Aware Speech Enhancement using Deep Neural Networks}
\name{Bahareh Tolooshams$^{\star}$ \qquad Kazuhito Koishida$^{\dagger}$
\thanks{Work done while B. Tolooshams was a Research Intern at Microsoft.}}
\address{
$^{\star}$School of Engineering and Applied Sciences, Harvard University, Cambridge, MA\\
$^{\dagger}$ Microsoft Corporation, One Microsoft Way, Redmond, WA\\
\vspace{5mm}
\href{mailto:btolooshams@seas.harvard.edu}{$\mathrm{btolooshams@seas.harvard.edu}$}, \href{mailto:kazukoi@microsoft.com}{$\mathrm{kazukoi@microsoft.com}$}
}

\usepackage{colortbl}
\definecolor{subjcolor}{RGB}{255,255,255}
\newcolumntype{s}{>{\columncolor{subjcolor}}c}
\definecolor{test1color}{RGB}{255,255,255}
\newcolumntype{t}{>{\columncolor{test1color}}c}

\begin{document}
\ninept

\maketitle

\begin{abstract}
Deep learning-based speech enhancement has shown unprecedented performance in recent years. The most popular mono speech enhancement frameworks are end-to-end networks mapping the noisy mixture into an estimate of the clean speech. With growing computational power and availability of multichannel microphone recordings, prior work has aimed to incorporate spatial statistics along with spectral information to boost up performance. Despite an improvement in enhancement performance of mono output, the spatial image preservation and subjective evaluations have not gained much attention in the literature. This paper proposes a novel stereo-aware framework for speech enhancement, i.e., a training loss for deep learning-based speech enhancement to preserve the spatial image while enhancing the stereo mixture. The proposed framework is model independent, hence it can be applied to any deep learning based architecture. We provide an extensive objective and subjective evaluation of the trained models through a listening test. We show that by regularizing for an image preservation loss, the overall performance is improved, and the stereo aspect of the speech is better preserved.
\end{abstract}

%%%%%%%%%%%%%%%%%%%%%%%%%%%%%%%%%%%%%%%
%%%%%%%%%%%%%%%%%%%%%%%%%%%%%%%%%%%%%%%
\section{Introduction}\label{sec:intro}

We consider the problem of stereo speech enhancement that is estimating a stereo clean speech from stereo noisy records. There exists a rich literature in signal processing for mono speech enhancement. To name a few, \citeauthor{ephraim1985lsa} enhance the speech through estimating its log-spectral amplitude~\cite{ephraim1985lsa}. \citeauthor{oppenheim1979enh} discuss various enhancement methods such as Wiener filtering and all-pole speech modeling~\cite{oppenheim1979enh}. Over the past decade, deep learning has gained a lot of attention for speech enhancement~\cite{wang2018supervised}; this is partly due to the growing number of available training datasets (i.e., clean speech and its noisy counterpart), and partly due to the outperformance of learning based approaches compared to classical methods~\cite{kaz2020low, soni2018tfgan, weninger2015speech}.

Learning based mono speech enhancement is mainly of two forms: a) predicting the clean speech through a deep neural network such as U-Net~\cite{ronneberger2015u}, or b) estimating a real~\cite{chakrabarty2018time, li2019tflstm} or complex~\cite{williamson2015complex, tolooshams2020ch} time-frequency (TF) mask such that when applied to the mixture it predicts the target speech. In the case of multichannel speech enhancement~\cite{li2019tflstm, tolooshams2020ch, gu2019end}, prior work focuses on extracting spatial features either explicitly at the input or implicitly through the network. \citeauthor{wang2018all} combine spectral features estimated through monaural speech enhancement with directional features to improve performance. \citeauthor{tolooshams2020ch} capture the spatial info with a beamforming-inspired architecture to perform complex ratio masking.

Despite the usage of spatial information within the network for speech enhancement, the preservation of spatial image such as sound image locations and sensations of depth is barely studied. Prior work on stereo enhancement mainly provides overall objective evaluations through metrics such as PESQ~\cite{rix2001pesq} and STOI~\cite{taal2011stoi} rather than focusing on perceptual enhancement through subjective tests. Moreover, in cases of reported subjective evaluations, mainly overall performance is studied rather than sound image~\cite{subramanian2019speechmushra, braithwaite2019speech, polyak2021regen}. We note that spatial cue preservation are studied previously for source separation~\cite{han2020mimotasnet}.

To fill the gap, this paper proposes a framework to preserve the stereo image and provide subjective evaluation along with objective metrics to assess the method. The approach is model-independent and fully focuses on training through a stereo-aware loss function helping to preserve spatial information. Specifically, the method regularizes to preserve interchannel intensity difference (IID), interchannel phase difference (IPD), interchannel coherence (IC), and overall phase difference (OPD). This is inspired by traditional methods~\cite{faller2003binaural, herre2004joint, breebaart2005parametric}, specifically parametric coding of stereo~\cite{breebaart2005parametric}, originally developed for efficient stereo coding to reduce bit-rate.
% breebaart2008spatial

\Cref{sec:methods} formulates the problem, introduces the stereo-aware training, and demonstrates the network architecture. The dataset, training details and evaluation metrics are explained in \Cref{sec:experiments}. We show in \Cref{sec:results} that the stereo-aware training not only results in an overall improvement of the enhanced speech, but also refines the stereo image. This is supported by both objective and subjective evaluation. Finally, \Cref{sec:conclusion} concludes.
%
%
%%%%%%%%%%%%%%%%%%%%%%%%%%%%%%%%%%%%%%%
%%%%%%%%%%%%%%%%%%%%%%%%%%%%%%%%%%%%%%%
\vspace{-3mm}
\section{Methods}\label{sec:methods}
\vspace{-3mm}
\subsection{Problem formulation}
Consider the discrete-time noisy speech $\y = [\y_1, \y_2] \in \R^{N \times 2}$ observed at a stereo microphone. In the stereo speech enhancement problem, we aim to estimate the clean reverberated stereo speech $\s \in \R^{N \times 2}$ given the mixture following the model
\vspace{-2mm}
\begin{equation}\label{eq:gen}
    \y[k] = \s[k] + \n[k]
\end{equation}
for $k = 1, \ldots, N$. The received speech $\s$ at the microphone is the result of convolving the speaker speech with stereo room impulse responses (RIRs). Similarly, the noise is reverberated through the room and recorded at the microphone as $\n$. In the time-frequency domain, the Short-Time Fourier Transform (STFT) of the mixture and speech are denoted as $\Y = [\Y_1, \Y_2]\in \C^{T \times F \times 2}$ and $\Source = [\Source_1, \Source_2]\in \C^{T \times F \times 2}$ with $T$ time frames and $F$ frequency bins.
\vspace{-2mm}
\subsection{Stereo-aware training}\label{sec:stereoloss}
Given the model-independence of the proposed framework, we focus mainly on the training loss enabling stereo image preservation and discuss the choice of neural architecture in \Cref{sec:network}. Given a set of training data, a network is trained to minimize a combination of \emph{speech reconstruction} and \emph{stereo image preservation} loss, i.e.,
\vspace{-2mm}
\begin{equation}\label{eq:loss}
 \mathcal{L}(\s, \hat \s) = \mathcal{L}_{\text{speech-rec}}(\s, \hat \s) + \mathcal{L}_{\text{image-pres}}(\s, \hat \s)
\end{equation}
where $\hat \s$ is an estimate of clean speech $\s$ given the mixture $\y$. We design the \emph{speech reconstruction} loss to suppress the noise and improve signal-to-noise ratio (SNR), and the \emph{image preservation} loss to conserve features related to the position of the speaker and microphone.

\textbf{Speech reconstruction:} Given $\s$ and $\hat \s$, the loss consists of log-spectral distortion (LSD)~\cite{wu2008minimum} and time loss (TL):
\vspace{-2mm}
\begin{equation}\label{eq:rec}
    \mathcal{L}_{\text{speech-rec}}(\s, \hat \s) = \text{LSD}(\s, \hat \s) + \alpha_{\text{TL}}\ \text{TL}(\s, \hat \s)
\end{equation}
with
\vspace{-2mm}
\begin{equation}\label{eq:lsd}
    \text{LSD}(\s, \hat \s) = \frac{1}{2T} \sum_{c=1}^2 \sum_{t=1}^{T} \sqrt{\frac{1}{F} \sum_{f=1}^F \left(g(\Source_c[t,f]) - g(\hat \Source_c[t,f])\right)^2}
\end{equation}
and 
\begin{equation}\label{eq:tl}
    \text{TL}(\s, \hat \s) = \frac{1}{2} \sum_{c=1}^2 \sqrt{\frac{1}{T} \sum_{t=1}^T \left(\s_c[t] - \hat \s_c[t] \right)^2}
\end{equation}
where $g(\x)$ is the generalized logarithmic function with $\gamma = \nicefrac{1}{3}$~\cite{kobayashi1984gen}. LSD helps to minimize the spectral error, and TL compensates for phase enhancement in the time domain.

\textbf{Image preservation:} We study the stereo image of the signal based on four spatial properties. We follow a similar approach to~\cite{breebaart2005parametric} and quantify the interchannel intensity-phase-coherence differences and overall phase. Although we study the parameters for image preservation, the original idea behind it is to reduce the bit-rate of the audio for a more efficient transmission or storage. For example, instead of transmitting the stereo signal, one may encode it with a mono downmix and stereo parameters. Then, the parameters are used by the decoder to reinstate spatial cues to reconstruct stereo~\cite{breebaart2005parametric}.

Given STFT $\Source_c = [\Source_c[1], \Source_c[2], \ldots, \Source_c[F]]$ for $c=1,2$, the frequency bins are grouped into $B$ non-overlapping subbands such that there are total of $32$ bins in each band. We leave non-uniform bands with equivalent rectangular bandwidth (ERB)~\cite{glasberg1990derivation} for future works. For $F = 1024$, there would be $B=32$ bands. For each band $b \in [1,2,\ldots, B]$, we extract IID, IPD, IC, and OPD as follows:
\vspace{-2mm}
\begin{equation}\label{eq:iid}
    \text{IID}_b(\Source) = 10 \log_{10}{ \frac{\sum_{f=f_b}^{f_{b+1}-1} \Source_1[f] \Source_1^*[f]}{\sum_{f=f_b}^{f_{b+1}-1} \Source_2[f] \Source_2^*[f]}}
\end{equation}
\vspace{-2mm}
\begin{equation}\label{eq:ipd}
    \text{IPD}_b(\Source) = \angle{\left( \sum_{f=f_b}^{f_{b+1}-1} \Source_1[f] \Source_2^*[f]\right)}
\end{equation}
\vspace{-2mm}
\begin{equation}\label{eq:ic}
    \text{IC}_b(\Source) = \frac{ |\sum_{f=f_b}^{f_{b+1}-1} \Source_1[f] \Source_2^*[f] |}{\sqrt{(\sum_{f=f_b}^{f_{b+1}-1} \Source_1[f] \Source_1^*[f]) (\sum_{f=f_b}^{f_{b+1}-1} \Source_2[f] \Source_2^*[f])}}
\end{equation}
\vspace{-2mm}
\begin{equation}\label{eq:opd}
    \text{OPD}_b(\Source, \hat \Source) = \angle{ \left(\sum_{f=f_b}^{f_{b+1}-1} \Source[f] \hat \Source^*[f]\right)}
\end{equation}
where we denote the frequencies in band $b$ by $[f_b, f_{b+1})$ and $*$ denotes complex conjugation. While IID and interchannel time differences cues are known to be useful for evaluation of sound source localization~\cite{rayleigh1907xii, sayers1964acoustic, bogaert2007wiener}, they have not been used during network training. We capture the time difference through IPD highlighting the delay between the channels. IC quantifies the correlation between the left and right channels given an aligned phase. Finally, OPD encodes the phase difference between the source and its estimate. Given the spatial parameters, image preservation error is defined as:
\begin{equation}\label{eq:stereo}
    \mathcal{L}_{\text{image-pres}}(\s, \hat \s) = \sum_{\text{M} \in \{\text{IID}, \text{IPD}, \text{IC}, \text{OPD} \}} \alpha_{\text{M}} \mathcal{L}_{\text{M}}(\Source, \hat \Source)
\end{equation}
\vspace{-4mm}
where
\begin{equation}\label{eq:iiderr}
    \mathcal{L}_{\text{M} \in \{\text{IID}, \text{IPD}, \text{ID}\}}(\Source, \hat \Source) = \frac{1}{T}\sum_{t=1}^{T} \sqrt{\frac{1}{B} \sum_{b=1}^B \left(\text{M}_b(\Source) - \text{M}_b(\hat \Source)\right)^2}
\end{equation}
\vspace{-2mm}
and
\begin{equation}\label{eq:opderr}
    \mathcal{L}_{\text{OPD}}(\Source, \hat \Source) = \frac{1}{2T} \sum_{c=1}^2 \sum_{t=1}^{T} \sqrt{\frac{1}{B} \sum_{b=1}^B \left(\text{OPD}_b(\Source_c, \hat \Source_c)\right)^2}
\end{equation}
\vspace{-2mm}
\begin{figure*}[t]
	\centering
	\begin{subfigure}[b]{0.999\textwidth}
	\includegraphics[width=0.999\linewidth]{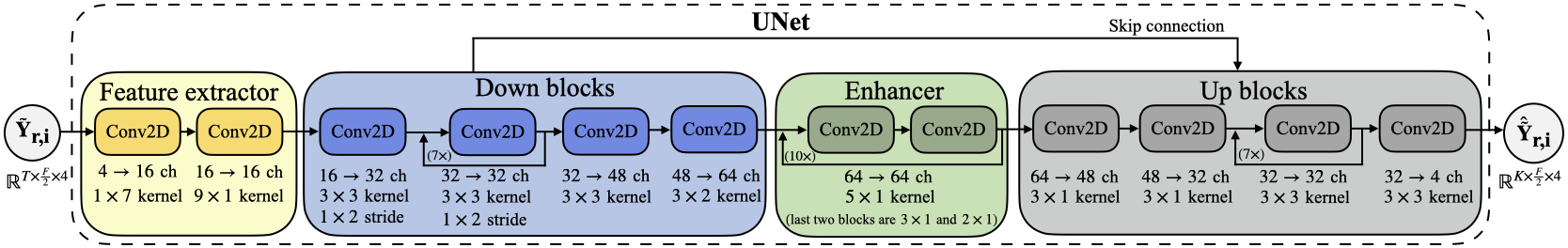}
 	 \caption{}
  	\label{fig:unet}
  	\end{subfigure}
	\centering
	\begin{subfigure}[b]{0.49\textwidth}
	\includegraphics[width=0.999\linewidth]{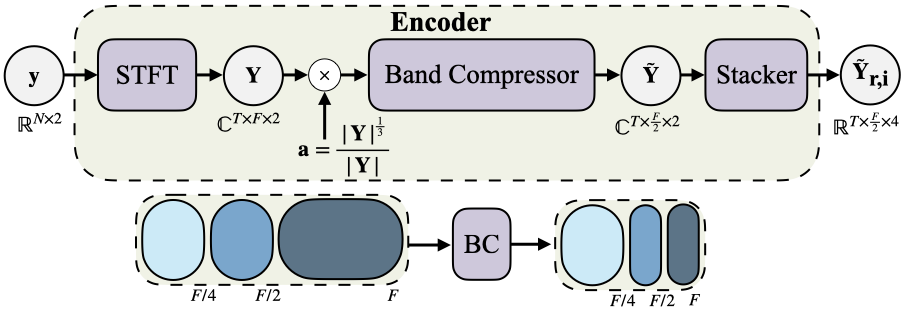}
 	 \caption{}
  	\label{fig:enc}
  	\end{subfigure}
	\centering
	\begin{subfigure}[b]{0.49\textwidth}
		\centering
% 		\vspace{-30mm}
	\includegraphics[width=0.999\linewidth]{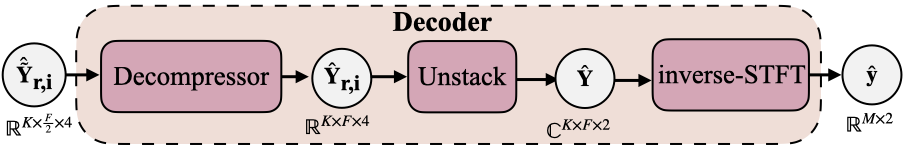}
 	 \caption{}
  	\label{fig:dec}
  	\end{subfigure}
  	\vspace{-4mm}
  	\caption{Network architecture. (a) U-Net denoiser. (b) Encoder. (c) Decoder.}
  	\label{fig:network}
  	\vspace{-2mm}
  	\vspace{-2mm}
\end{figure*}
%
%
%%%%%%%%%%%%%%%%%%%%%%%%
\vspace{-4mm}
\subsection{Network architecture}\label{sec:network}
%
% encoder
The network (\Cref{fig:network}) consists of three main blocks, i.e., encoder, denoiser, and decoder. Given the noisy input $\y$, the encoder computes its STFT $\Y$ and scales it by ${\bf a}$. Then, the signal is passed through a band compressor (BC) and is outputted as a stack of the real and imaginary components $ \tilde{\Y}_{r,i}$. BC compresses the mixture by a factor of $2$ along the frequency domain. Precisely, it passes the low frequency bins in $[0, \nicefrac{F}{4}]$, and compresses the bins in $[\nicefrac{F}{4}, \nicefrac{F}{2}]$ and high frequencies in $[\nicefrac{F}{2}, F]$ by a factor of $2$ and $4$, respectively. This compression is achieved by averaging the neighbouring frequencies.

% denoiser
The denoiser has a U-Net structure~\cite{ronneberger2015u} consisting of a feature extractor ($2$ blocks), down-blocks ($11$ blocks), enhancer ($10$ blocks), and up-blocks ($10$ blocks). The architecture has skip-connections between down and up-blocks. The building blocks contain convolution layers, causal along the time axis. All blocks have leaky ReLU activations with $\alpha = 0.2$ (except the first extractor block) and batch normalization (except last up-block). The up-blocks contain pixel shufflers to reshape the feature map into desired number of channels.

The decoder decompresses the signal to reverse the BC operation, and applies an inverse STFT to construct the signal in time domain. The main results are based on this U-Net architecture. To emphasize the model-independence of the stereo-aware framework, we additionally train a similar architecture, which we call U-NetCM, with a decoder estimating a complex TF mask for enhancement~\cite{williamson2015complex, tolooshams2020ch}.
%
%
%%%%%%%%%%%%%%%%%%%%%%%%%%%%%%%%%%%%%%%
%%%%%%%%%%%%%%%%%%%%%%%%%%%%%%%%%%%%%%%
\vspace{-2mm}
\section{Experiments}\label{sec:experiments}
\vspace{-2mm}
\subsection{Dataset}\label{sec:dataset}
Both training and testing data are sampled at $48$ kHz. We use the Deep Noise Suppression (DNS) challenge dataset~\cite{reddy2021interspeech} to generate training stereo data. We picked mono clean and noise tracks at random, and applied RIRs to create stereo (usage of RIRs instead of head-related transfer function is to focus on stereo image on the recording device). Then, clean and noise are mixed with an SNR sampled from $\mathcal{N}(5, 100)$ with a range of $[-10, 30]$ dB. The signals are leveled up/down using a scale following $\mathcal{N}(-26, 100)$. We generate approximately $1.086$ M stereo signals. During training, a random segment of $1.94$ s is selected, i.e., $N=93{,}120$ samples.
% training data 1,086,282

We use two test sets. For each, we create $560$ mono utterances, and construct a stereo test set using test RIRs. The sets are divided into five groups each with SNR of $0$, $5$, $10$, $15$, and $20$ dB. The speech and mixture are scaled by a constant following $\mathcal{N}(-2, 4)$.

\textbf{Room impulse responses (RIR):} RIRs are simulated using an image method similar to~\cite{rao2021interspeech}. The room size ranges from $3 \times 3 \times 8$ to $8 \times 8 \times 8$ $\text{m}^3$. The microphones are $20$ cm apart and are placed in the room uniformly at random such that their height ranges from $1$ to $1.5$ m, and they are within the second and third quarter in the middle. The speaker and noise are randomly located in the room with a height ranged from $1.2$ to $1.9$ m. The speaker and noise have distances of $[0.5, 2]$ m and $[1.5, 2]$ m from the microphone, respectively. Finally, we make sure that the angle between the speech and noise is at least $20^{\circ}$, and sound velocity is $340$ m/s. The generated RIRs are $0.9$ s long and categorize into two groups of with and without reverberation.

We create around $4{,}640$ training rooms and generate $10$ utterances for each. For the training set, there are $12{,}000$ no reverberation RIRs, and $34{,}400$ reverberated RIRs with $60$ dB attenuation time sampled from $\text{Unif}[0.2, 0.8]$ s. For each test set, $560$ rooms (i.e., one for each test example) are created. The sets follow similar characteristics as in the training set, except that Test set II uses a room height of $3$ m (i.e., a typical meeting room). The test set RIRs are divided into four categories of no, short, medium, and long reverberation which has $60$ dB attenuation in $0, 0.27, 0.53, 0.8$ s, respectively.
%
% rir reverb 34,415, rir no reverb 12,000
%
%
\vspace{-2mm}
\subsection{Training}\label{sec:training}
The network is trained using ADAM optimizer with $\beta_1 = 0.5$, $\beta_2 = 0.9$, and an initial learning rate of $10^{-4}$. The learning rate is scheduled with piecewise constant decay to $10^{-5}$ after $300{,}000$ iterations. Training is performed on four GPUs using a batch size of $64$ for $350{,}000$ iterations. For TL, $\alpha_{\text{TL}}$ is set to $50$. Additionally, $\alpha_{\text{IID}} = 0.05$, $\alpha_{\text{IPD}} = 0.05$, $\alpha_{\text{IC}} = 0.4$, and $\alpha_{\text{OPD}} = 0.05$ whenever the particular error is present. The above weights are chosen such that all loss components are on the same order. STFT and inverse-STFT blocks use Hanning windows of length $2048$ (i.e., $F = 1024$) with hop size of $480$. Then, $T=192$ time frames are cropped. 
\vspace{-3mm}
\subsection{Evaluation}\label{sec:evaluation}
Signal-to-distortion ratio (SDR) and perceptual objective listening quality assessment (POLQA)~\cite{beerends2013perceptual} along with stereo preservation errors are used as objective metrics. Given the enhanced speech, SDR and POLQA metrics (higher the better) are computed independently for each channel, and the average is reported. Additionally, we quantify the errors for IID, IPD, IC and OPD (lower the better).

We perform a listening test following MUSHRA standards~\cite{series2014method} through a vendor specialized in designing cloud-based tests. Approximately $2{,}750$ listeners, wearing headphones, participated in the experiments; each evaluates a subset of the test set given OVRL or IMG task. Given a reference (i.e., clean speech), listeners are ask to evaluate and grade several tracks including a hidden reference and an anchor, i.e., the noisy mixture~\cite{polyak2021regen, deng2015speech, braithwaite2019speech}.
We evaluate two attributes (OVRL and IMG). For OVRL, the assessors are asked to evaluate the overall quality of the audio clips. For IMG, they rate the stereophonic image quality of the clips (i.e., how close the clips are to the reference in terms of sound image locations, sensations of depth, and reality of the speaker). We categorize the MUSHRA grading scheme from $1$ to $5$ as ($1$) Bad, ($2$) Poor, ($3$) Fair, ($4$) Good, and  ($5$) Excellent.

\begin{table*}[!t]
  \renewcommand{\arraystretch}{0.9}
  \centering
  \caption{Evaluation results on stereo test sets.}
  \setlength\tabcolsep{2.pt}
  \centering
  \begin{tabular}{c|c||tttttt|ss||cccccc}
    \bottomrule
    \multirow{3}{*}{Network} & \multirow{3}{*}{Method} & \multicolumn{8}{c||}{Test set I} & \multicolumn{6}{c}{Test set II}\\
    \cline{3-16}
    & & \multicolumn{6}{c|}{Objective} & \multicolumn{2}{c||}{Subjective} & \multicolumn{6}{c}{Objective}\\
    \cline{3-16}
    & & SDR & POLQA & IID & IPD & IC & OPD & OVRL & IMG & SDR & POLQA & IID & IPD & IC & OPD\\
    \hline \hline
    & \textit{noisy} & 11.61 & 2.51 & 1.56 & 1.92 & 0.20 & 0.78 & 0 & 0 & 11.13 & 2.50 & 1.60 & 1.96 & 0.18 & 0.79\\
    \hline
    \multirow{10}{*}{U-Net} & \textit{downmix - spec} & 6.46 & 2.98 & 2.68 & 2.79 & 0.30 & 1.61 & x & x & 6.16 & 2.95 & 2.70 & 2.83 & 0.31 & 1.62 \\
    & \textit{LRindp - spec} & 6.82 & 3.26 & 2.36 & 1.99 & 0.28 & 1.62 & x & x & 6.67 & 3.19 & 2.48 & 2.02 & 0.27 & 1.63 \\ 
    % \textit{stereo - spec} & 2.99 & \bf 3.42 & 1.68 & 2.48 & 0.40 & 1.90 & - & - & 2.36 & 3.32 & 1.71 & 2.49 & 0.41 & 1.90 \\
    & \textit{downmix - spec - time} & 10.10 & 2.95 & 2.39 & 2.78 & 0.29 & 1.40 & 0.34 & 0.30 & 9.65 & 2.92 & 2.42 & 2.82 & 0.29 & 1.40 \\ 
    & \textit{LRindp - spec - time}  & 12.89 & 3.31 & 2.42 & 1.92 & 0.27 & 1.27 & 0.42 & 0.35 & 12.27 & 3.24 & 2.55 & 1.95 & 0.26 & 1.27 \\
    % \hline
    & \textit{stereo - spec - time} & 12.56 & 3.01 & 1.85 & 1.91 & 0.26 & 1.25 & 0.38 & 0.37 & 11.97 & 2.96 & 1.90 & 1.93 & 0.28 & 1.23 \\
    & \textit{stereo - spec - time - IID} & \bf 14.17 & 3.33 & \bf 1.55 & 1.76 & 0.35 & 1.42 & 0.45 & 0.41 & \bf 13.64 & 3.26 & \bf 1.59 & 1.79 & 0.39 & 1.43 \\
    & \textit{stereo - spec - time - IPD} & 13.88 & \bf 3.36 & 1.67 & \bf 1.71 & 0.32 & 1.27 & \bf 0.63 & 0.46 & 13.24 & \bf 3.30 & 1.71 & \bf 1.73 & 0.36 & 1.28 \\
    & \textit{stereo - spec - time - IC} & 12.09 & 3.04 & 1.80 & 2.08 & \bf 0.21 & 1.43 & 0.31 & 0.37 & 11.47 & 2.98 & 1.85 & 2.12 & \bf 0.20 & 1.40 \\
    & \textit{stereo - spec - time - OPD} & 14.05 & 3.33 & 1.86 & 2.10 & 0.23 & \bf 0.99 & 0.42 & \bf 0.49 & 13.35 & 3.28 & 1.90 & 2.15 & 0.22 & \bf 1.00 \\ 
    & \textit{stereo - spec - time - all} & 13.78 & 3.32 & 1.64 & 1.81 & \bf 0.21 & 1.10 & 0.45 & 0.43 & 13.16 & 3.25 & 1.69 & 1.85 & \bf 0.19 & 1.11 \\
    \hline %\hline
     \multirow{2}{*}{U-NetCM} & \textit{stereo - spec} & 6.28 & \bf 3.34 & 2.24 & 2.14 & 0.25 & 2.48 & x & x & 6.10 & \bf 3.27 & 2.29 & 2.18 & \bf 0.23 & 2.46 \\
    & \textit{stereo - spec - time - all} & \bf 15.02 & 3.28 & \bf 1.96 & \bf 1.93 & \bf 0.24 &  \bf 1.05 & x & x & \bf 14.30 & 3.22 & \bf 2.01 & \bf 1.97 & \bf 0.23 & \bf 1.06 \\
    \hline
    \bottomrule
  \end{tabular}
  \label{tab:res}
  \vspace{-4mm}
\end{table*}
%
%
%%%%%%%%%%%%%%%%%%%%%%%%%%%%%%%%%%%%%%%
%%%%%%%%%%%%%%%%%%%%%%%%%%%%%%%%%%%%%%%
\vspace{-2mm}
\vspace{-2mm}
\section{Results}\label{sec:results}
We train the stereo network using various combinations of the loss. We denote the presence of LSD and TL in the training loss by \textit{spec} and \textit{time}, respectively. For example, \textit{spec}-\textit{time}-\textit{OPD} denotes the case where loss includes LSD, TL, and OPD errors. Given the rich deep learning literature on enhancing mono signals, we consider two baselines. A mono network that is trained using downmix (i.e., $\nicefrac{(L + R)}{2})$, where for prediction, the phase difference between the mixture stereo and enhanced downmix is added to reinstate stereo. We call this method \textit{downmix}. The other baseline, \textit{LRindp}, is a mono network trained using left and right channels independently. We first compare the baselines to one another, then highlight the effect of the time loss on the performance, and finally focus on the stereo-aware training. \Cref{tab:res} demonstrates the evaluations on the test sets where the comparisons we highlight bellow holds for both test sets.

\textbf{Downmix vs. LRindp:} \textit{LRindp-spec} shows better performance in terms of SDR and POLQA against \textit{downmix-spec} and also results in a better image preservation (lower IID, IPD, and IC errors). In spite of the better performance, \textit{LRindp} has approximately doubled inference time compared to \textit{downmix}. Drawbacks of \textit{downmix} may come from the addition of noisy phase at channel-upsampling time. 

\begin{figure}[t]
	\centering
	\begin{subfigure}[b]{0.48\linewidth}
	\includegraphics[width=0.98\linewidth]{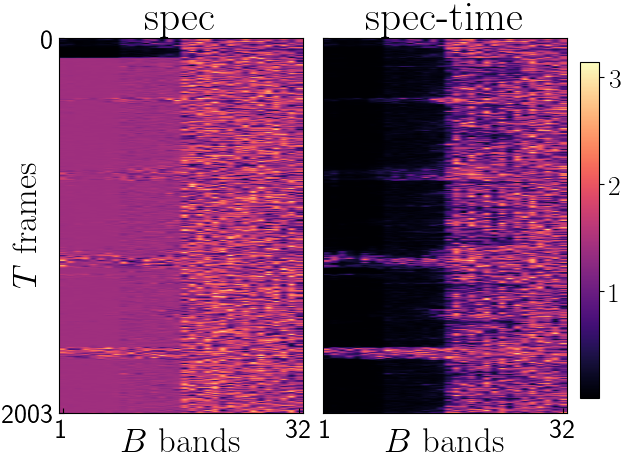}
	\vspace{-2mm}
  	\caption{$|\text{OPD}|$ for $LRindp$ ($\text{c}=1)$.}
  	\label{fig:timeloss}
  	\end{subfigure}
 	\centering
 	\begin{subfigure}[b]{0.48\linewidth}
	\includegraphics[width=0.98\linewidth]{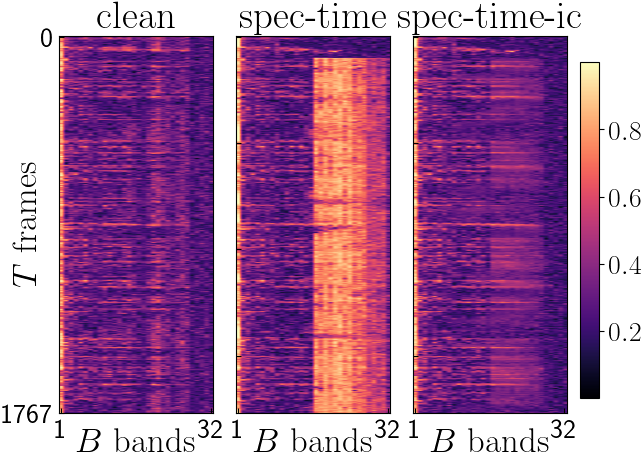}
	\vspace{-2mm}
  	\caption{IC preservation.}
  	\label{fig:ic}
  	\end{subfigure}
 	\centering
	\begin{subfigure}[b]{0.47\linewidth}
	\includegraphics[width=0.98\linewidth]{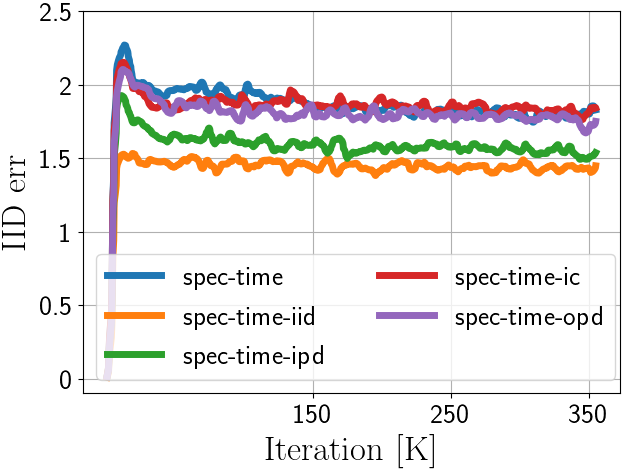}
	\vspace{-2mm}
  	\caption{IID training dynamics.}
  	\label{fig:iiddyn}
  	\end{subfigure}
 	\centering
	\begin{subfigure}[b]{0.47\linewidth}
	\includegraphics[width=0.98\linewidth]{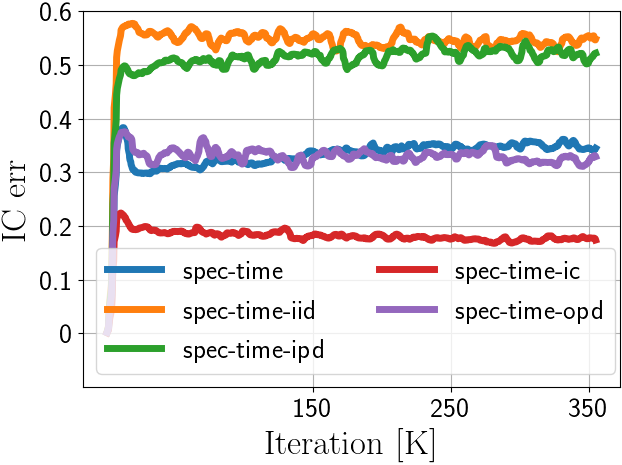}
	\vspace{-2mm}
  	\caption{IC training dynamics.}
  	\label{fig:icdyn}
  	\end{subfigure}
  	\vspace{-2mm}
  	\caption{Visualization of stereo-parameters and errors.}
  	\vspace{-2mm}
  	\vspace{-4mm}
\end{figure}

\textbf{Presence of time loss:} Comparing \textit{spec} with \textit{spec-time}, time loss results in a drastic improvement in SDR with a trade-off being an occasional decrease in POLQA. TL also helps to preserve OPD. \Cref{fig:timeloss} highlights the overall phase preservation using the time loss; compared to \textit{LRindp-spec}, \textit{LRindp-spec-time} has lower OPD in magnitude, particularly at low-frequency bands.

\textbf{Mono to stereo:} Moving from mono to stereo, we observe that the stereo image of \textit{LRindp-spec-time} gets worse, but that of the \textit{stereo-spec-time} is improved. Stereo method preserves IID much better than the mono case. However, \textit{LRindp-spec-time} (with doubled inference time) has better performance in terms of SDR and POLQA compared to \textit{stereo-spec-time}. This motivates us to regularize for image preservation loss which cannot be done in a mono network. We show how this helps the \textit{stereo} network to outperform \textit{LRindp}.

\textbf{Image dynamics during training:} We first study the effect of each stereo parameter independently. \Cref{fig:iiddyn} shows how the IID loss for the training batch changes as a function of training iterations; specifically, it shows that IPD regularization alone helps to achieve a lower IID error and the lowest IID is achieved when the IID loss is presented. Moreover, \Cref{fig:icdyn} demonstrates that regularizing for IID or IPD preservation results in a worse IC than no image regularization (i.e., \textit{stereo-spec-time}). The figure highlights that OPD does not have much effect on IC, and IC can further be improved by regularizing for IC. Finally, we observe (not shown) that including IC loss during training results in a worse IPD compared to no regularization training.

\textbf{Image preservation loss:} Given \textit{stereo-spec-time}, the addition of IID, IPD, IC, or OPD results in a POLQA improvement. Furthermore, regularizing for IID, IPD, or OPD, improves SDR. We observed that regularizing for the preservation of a stereo metric alone (e.g., \textit{stereo-spec-time-IC}) results in the best preservation of that metric (e.g., \text{IC}) in the test sets among all other methods. \Cref{fig:ic} visualizes IC of the clean speech, predicted signal through \textit{stereo-spec-time} and \textit{stereo-spec-time-IC} from Test set I. The figure highlights that \textit{stereo-spec-time-IC} has preserved IC better than \textit{stereo-spec-time}. Results show that IID helps the best to improve SDR, and IPD results in the highest POLQA improvement. Overall, compared to the unregularized case, \textit{stereo-spec-time-all} preserves all aspects of the stereo image. Finally, we note the subjective results may contain uncertainties in evaluation of the stereo image as it is challenging to fully ignore the speech distortion while scoring the stereo image.

\textbf{Subjective evaluation:} We conduct four tests on Test set I. In each test, five tracks (i.e., three methods and hidden noisy and reference) are compared. To combine the results, we report the mean of relative score with respect to hidden noisy in each test (higher the better). The ``Subjective'' column of \Cref{tab:res} demonstrates the result of this listening test where the average relative difference of hidden reference and noisy is $1.19$ and $1.15$ for OVRL and IMG, respectively. The table demonstrates that \textit{stereo-spec-time-IPD} achieves the highest OVRL score which also has highest POLQA among all. Moreover, all stereo-aware training methods results in higher IMG ($0.49$ at highest) score compared to \textit{downmix} ($0.3$) and \textit{LRindp} ($0.35$). We emphasize that the inference complexity of our stereo networks is approximately half of \textit{LRindp}. Among the stereo-aware regularization methods, the listeners have given highest IMG scores of $0.49$ and $0.46$ when OPD and IPD, respectively, are preserved the best (i.e., \textit{stereo-spec-time-OPD/IPD}). These results highlight the benefits of the proposed training approach in subjectively improving the image.

\textbf{Model independence:} We lastly apply a stereo-aware training framework on a different architecture, U-NetCM (last row of ~\Cref{tab:res}). We observe that including all the stereo errors along with time results in lower stereo image errors and an improvement in SDR.

%
%
%%%%%%%%%%%%%%%%%%%%%%%%%%%%%%%%%%%%%%%
%%%%%%%%%%%%%%%%%%%%%%%%%%%%%%%%%%%%%%%
\vspace{-3mm}
\section{Conclusion}\label{sec:conclusion}
This paper studied the perceptual enhancement of stereo speech. The paper proposed a stereo-aware training loss to preserve the image while aiming to estimate the clean speech from noisy mixture. The trained architecture was a variant of a causal U-Net and the image preservation loss consist of errors related to interchannel intensity and phase differences, interchannel coherence, and overall phase. We showed that accounting for preservation of the stereo image improves the enhancement both objectively with the SDR and POLQA metrics and subjectively through a MUSHRA listening test.
%
%
%
%
%
%%%%%%%%%%%%%%%%%%%%%%%%%%%%%%%%%%%%%%%
%%%%%%%%%%%%%%%%%%%%%%%%%%%%%%%%%%%%%%%
% references
\bibliographystyle{IEEEtranN}
\bibliography{2022-icassp-stereo}
\end{document}